\documentclass[aps, prl,twocolumn,groupedaddress, floatfix,superscriptaddress]{revtex4-1}
\usepackage{graphicx}
\usepackage{transparent}
\usepackage{bm}
\usepackage{amsmath}
\usepackage{natbib}
\usepackage{color}
\usepackage{epstopdf}
\begin{document}
\title{Stirrers and movers actuated by oscillating fields}

\author{Gabi Steinbach}
\affiliation{Helmholtz-Zentrum Dresden-Rossendorf, Institute of Ion Beam Physics and Materials Research, Bautzner Landstrasse 400, 01328 Dresden, Germany.}
\affiliation{Institute of Physics, Technische Universit\"at Chemnitz, 09107 Chemnitz, Germany.}
\author{Michael Schreiber}
\affiliation{Institute of Physics, Technische Universit\"at Chemnitz, 09107 Chemnitz, Germany.}
\author{Dennis Nissen}
\affiliation{Institute of Physics, University of Augsburg, 86159 Augsburg, Germany.}
\author{Manfred Albrecht}
\affiliation{Institute of Physics, University of Augsburg, 86159 Augsburg, Germany.}
\author{Sibylle Gemming}
\affiliation{Institute of Physics, Technische Universit\"at Chemnitz, 09107 Chemnitz, Germany.}
\affiliation{Helmholtz-Zentrum Dresden-Rossendorf, Institute of Ion Beam Physics and Materials Research, Bautzner Landstrasse 400, 01328 Dresden, Germany.}
\author{Artur Erbe}
\affiliation{Helmholtz-Zentrum Dresden-Rossendorf, Institute of Ion Beam Physics and Materials Research, Bautzner Landstrasse 400, 01328 Dresden, Germany.}

\begin{abstract}

Locomotion via cyclic moves presents a challenge to mesoscopic objects in overdamped environments, where time reversibility may prevent directed motion. Most reported  cyclic movers exploit anisotropic drag to push themselves forward. Under an oscillating drive, however, anisotropic drag enables locomotion only if the objects can change their shape. Here, we present a strategy that unexpectedly enables structurally invariant objects to move under oscillating fields. The objects are self-assembled clusters of magnetic particles that exhibit an off-centered dipole moment. By theoretical modeling and in experiments with magnetic Janus particles, we demonstrate that the interaction between such anisotropic particles in the cluster breaks time reversibility.  Experimentally, we show that the magnetic configuration of a cluster determines its motion path. We realize stirrers and steerable movers with helical or directed path using the same particle system. The presented strategy based on internal interactions establishes a counterpart to locomotion via anisotropic drag.
\end{abstract}

\maketitle



Since  Purcell proposed the Scallop theorem in 1977 \cite{Pur77}, locomotion of microscopic objects has been the subject of intense research \cite{Lau09,Bec16,Elg15,Gol11, Dun09}. The fascination for migration at this length scale arises from the time reversibility of motion patterns at low Reynolds numbers, where drag dominates over inertia. 
 Under any oscillating pattern of motion, a microscopic object repeatedly goes back and forth by the same distance. Such a so-called reciprocal motion excludes directed migration. 
Yet, nature has found ways for microorganisms to move  \cite{Fle09,Sto96,Arr12,Tam11,Gol15,Wan16,Elg13,Jun14}. 
 Typically, these organisms exploit anisotropic drag under cyclic body deformations to break time reversibility. Understanding propulsion strategies at low Reynolds numbers is crucial in exploring microbial life. 
Artificial systems have been established - using field-activated or catalytically active objects \cite{Abb09,Rao15,Zot16,Naj04,Pal16,Pal13,Erb08,Vac15} - to study the principles of locomotion.  Moreover, they provide practical use as artificial micromotors for medical and lab-on-chip applications \cite{Cha15,Ree14,Ebb16}.
 
For locomotion via cyclic moves, magnetic particle dispersions are versatile systems \cite{Vac15,Fis11,Kha15,Mar15,Tie14,Erb16}.  Under rotating or oscillating magnetic fields, ferromagnetic particles periodically rotate via alignment in the field. The challenge is to  transform field-driven rotation  by an angle $\theta$ into an effective translation $r$.   
This can be realized only if, first, the rotation creates a force on the center of mass of the particle, and, second, the motion pattern is non-reciprocal. 

\begin{figure}[b]
\centerline{\includegraphics[width=8.7cm]{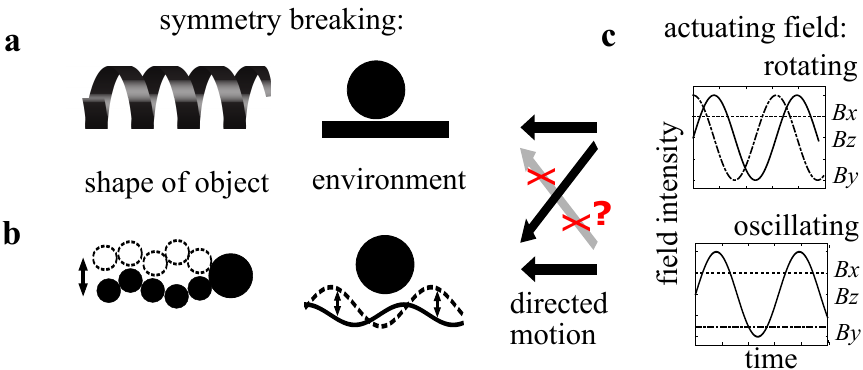}}
\caption{Overview  over magnetic actuation through anisotropic drag. Types of  symmetry breaking in (\textbf{a}) rigid systems and (\textbf{b}) flexible systems, and (\textbf{c}) actuating fields.}
\label{sktrans}
\end{figure}  
To date, most examples of magnetic actuation (and other cyclic movers) exploit hydrodynamic coupling between rotation and translation  \cite{Vac15,Tie14}. In rigid systems, coupling appears  if the object has a chiral shape  (propeller) \cite{Zha09,Cas13,Gho09},  or if the environment is anisotropic, e.g., close to a  surface (roller)  \cite{Gol67a,Tie08a,Zha10,Tie12} (Fig.\,\ref{sktrans}\,a). Introducing flexibility extends the possibilities (Fig.\,\ref{sktrans}\,b).  Multi-component objects with a flexible tail,  resembling flagella (swimmer) \cite{Dre05,Ben11,Gau06}, perform undulations under field alignment, and hydrodynamic drag leads to translation. Also, rigid objects can be actuated if they are close to  an elastic surface \cite{Tro08}. 
An essential difference between rigid and flexible systems is that both enable locomotion under rotating fields, but only flexible systems can be actuated under oscillating fields. In contrast, rigid systems perform a reversible back and forth motion between maximum rotation angles $\pm\theta_A$. 
Under time reversibility, locomotion can be achieved only if at least two coordinates of motion of the object are affected in a time-dependent manner. Trivially, this is the case under rotating  fields, which act along two coordinates (Fig.\,\ref{sktrans}\,c). 
 In contrast, an oscillating field exerts a torque along one coordinate only. There, flexibility provides an additional coordinate of motion that can break time reversibility. 
Studies to date consider flexibility of structural deformations only (Fig.\,\ref{sktrans}\,b). This disregards, however, that widely discussed multi-component objects (whether invariant \cite{Cas13} or flexible \cite{Dre05} in object shape) provide additional  flexibility when the components rotate inside the object. The problem is that internal rotations do not break time reversibility within the concept of  hydrodynamic coupling; an alternative strategy is required. 

Here, we present an actuation concept where internal rotations in multi-component objects cause propulsion  under oscillating fields even if the shape of the object remains constant.  
 We study self-assembled clusters of a few, rotatable spheres that exhibit an off-centered net dipole moment. The clusters themselves are structurally invariant since the relative distances between all spheres remain constant. 
 We explain the actuation concept qualitatively with a simple theoretical model. It assumes spheres with an embedded dipole that is not centered but shifted radially outwards, so-called shifted-dipole particles (sd-particles). 
The key point is that the interaction between spheres with this magnetic asymmetry breaks time-reversal symmetry under oscillating fields.
 The theoretical predictions are verified with an experimental system of magnetic Janus spheres. Experimentally, we realize a variety of motion paths, which depend on the cluster configuration. The findings demonstrate that internal rotations in multi-component objects enable new actuation strategies that complement established schemes based on anisotropic drag. 


\section{Results}
The actuation concept  consists of two elements. First, the conversion of rotation into time-reversible translation is achieved by each particle separately, hence, a single-particle problem. Second, the breaking of the time reversibility  arises from the interaction between the particles in a cluster. In the following, these two elements are demonstrated qualitatively with a combination of analytic and numerical calculations using sd-particles. Afterwards, the investigations are used to explain our experimental observations of actuated clusters from magnetic Janus spheres, which move under oscillating fields.

\subsection{Oscillation of single sd-particles} 
\begin{figure}
\centerline{\includegraphics[width=8.7cm]{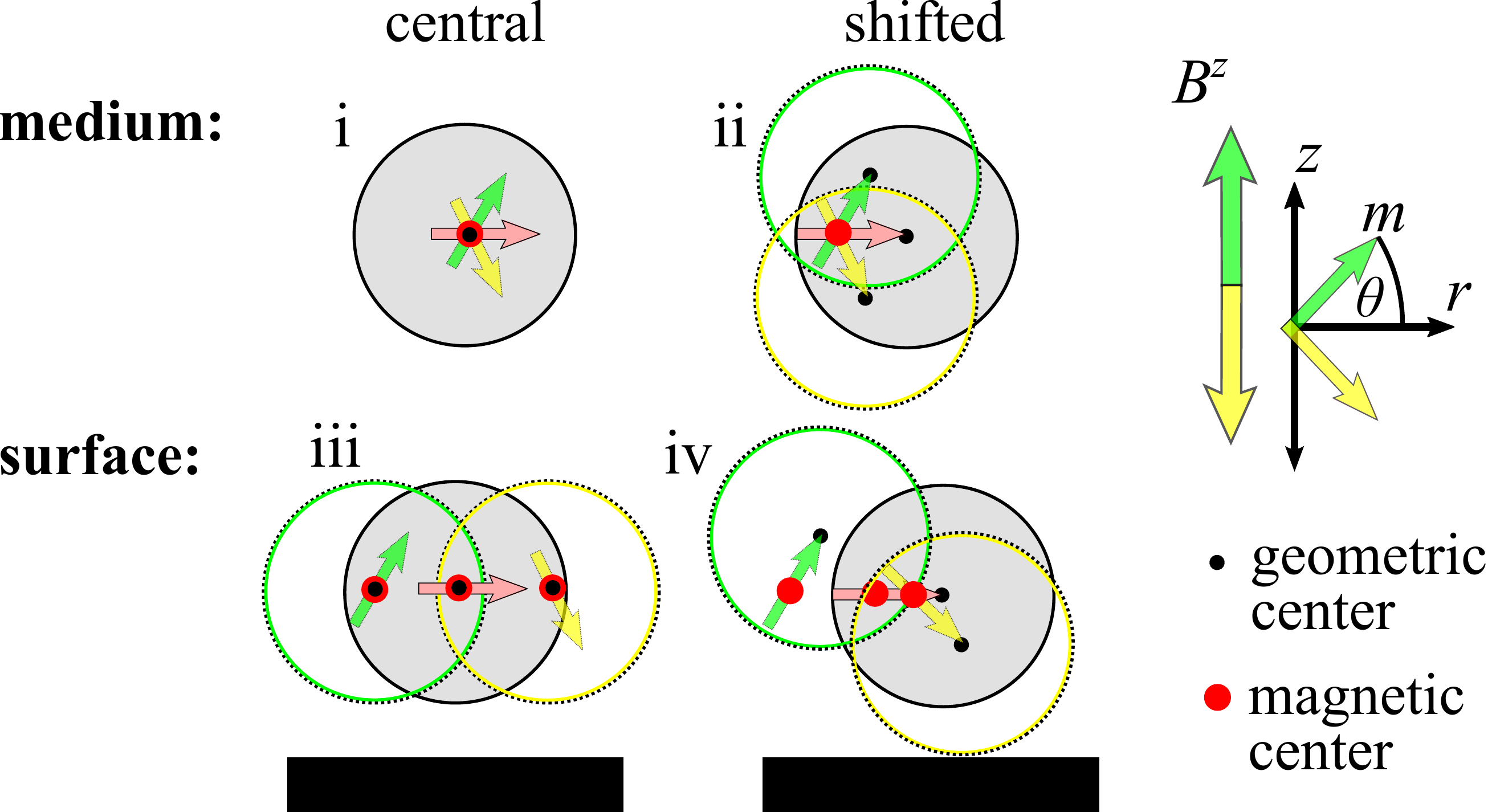} }
\caption{Time-reversible motion of a single dipolar sphere under oscillating magnetic fields $B^{\text{z}}$. The scenarios depict a sphere with central dipole (i, iii) and shifted dipole (ii, iv) in an isotropic medium (i, ii) and close to a wall (iii, iv). The dipole (light red arrow) with moment $m$  oscillates up (green) and down (yellow), which results in a displacement of the sphere (dotted green and yellow circles) under asymmetric conditions (ii\,-\,iv).
}
\label{pics:sk-trans-sd} 
\end{figure}

We access the role of magnetic asymmetry of a single dipolar particle by considering its reciprocal motion  under oscillating fields for four different scenarios (Fig.~\ref{pics:sk-trans-sd}). If, in an isotropic medium, a sphere with a central dipole (case i, Fig.~\ref{pics:sk-trans-sd}) is exposed to an oscillating field $B^{z}=B_0^{z}\sin(\omega_B t)$, the dipole follows the field via up  and down rotation by angle $\theta\in [-\theta_A,\theta_A]$. As a magnetic object rotates around its magnetic center this results only in particle rotation.
 Asymmetry is introduced by sd-particles, where the magnetic center is shifted away from the geometric center (case ii, Fig.~\ref{pics:sk-trans-sd}). Here, the dipole is shifted radially outwards  by $\xi\,\frac{d_p}{2}$ along the dipole orientation, where $\xi\in[0,1]$ is the shift parameter and $d_p$ is the particle diameter. Under oscillating fields, the rotation around the magnetic center displaces the particle center periodically. 
In an isotropic environment, the particle center moves on an arc \cite{Che09} with a radius determined by the shift $\xi$. 
The behavior changes in the proximity of a wall, where anisotropic drag is present. We consider this case here because the experiments described later in this work are performed on particles moving close to a planar substrate. We emphasize that the presence of a wall is, however, not essential for the actuation mechanism presented here.  

 Already for a sphere with central dipole (case iii, Fig.~\ref{pics:sk-trans-sd}) in a liquid medium close to a wall, rotation is converted into reversible back and forth displacement. An sd-particle  also rolls back and forth, and additionally performs a periodic displacement away from the wall and towards the wall (case iv, Fig.~\ref{pics:sk-trans-sd}) because the magnetic center is shifted away from the geometric center of the sphere.  In contrast to case iii, this motion is not symmetric with respect to  $\theta$ due to the hindrance by the wall. 

For an sd-particle in an isotropic medium, we can derive an analytic expression for the
 displacement of the particle center that arises from the rotation by an infinitesimal angle $\mathrm{d}\theta$ around the magnetic center (see SI Appendix). The infinitesimal displacement,  separated into components  perpendicular ($\mathrm{d}r$) and  parallel ($\mathrm{d}z$) to the oscillating field $B^{z}$, is given by 
 
\begin{subequations}
\begin{align}
\mathrm{d}r &= - \xi \frac{d_p}{2} c(\theta) \sin \theta\, \mathrm{d}\theta ,   \\
\mathrm{d}z &= \xi \frac{d_p}{2} c(\theta) \cos \theta\, \mathrm{d}\theta .
\end{align}
  \label{eq:conv-med}%
\end{subequations}
The  term $c(\theta)$ contains potential sources of anisotropic hindrance  or enhancement.
For a single particle, this is given by, e.g., a wall. The determination of $c(\theta)$ close to a wall is a complex problem. Since the rotation around the shifted magnetic center periodically changes the distance between particle center and wall, also the hydrodynamic friction coefficient changes periodically. To illustrate the concept, here, it is sufficient to approximate the surface-induced drag with a step function of two  values $c_1,c_2$ that alternate with the sign of $\theta$. Distinguishing  between translation parallel ($r$) and perpendicular ($z$) to the surface gives
\begin{equation}
c_r(\theta)= \begin{cases} 
   c_{r,1} & \text{if } \theta \geq 0 \\
   c_{r,2}       & \text{if } \theta  < 0
  \end{cases}
  \text{ ,}  \quad
  c_z(\theta)= \begin{cases} 
   c_{z,1} & \text{if } \theta \geq 0 \\
  c_{z,2}       & \text{if } \theta  < 0
  \end{cases}.
  \label{eq:conv-sf}
\end{equation}
During $\theta > 0$ (green circle in case iv, Fig.~\ref{pics:sk-trans-sd}) the surface-induced translation (ii) enhances the translation $r$ caused by the off-centered dipole (iii). During $\theta < 0$ (yellow circle), the directions of the two translations are opposite to each other as can be seen by the comparison between cases ii and iii. Therefore, for oscillations  close to a wall (iv) the displacement during  $\theta > 0$  is always higher than during $\theta < 0$ and, thus,  $|c_1|>|c_2|$. If the influence of the surface on the displacement is stronger than that of the shifted magnetic center, then  during $\theta<0$ the overall translation direction  is even inverted with respect to the one in an isotropic medium (ii), and $c_{r,2}<0$. Otherwise $c_{r,2}>0$ holds.  Note that due to the time reversibility the integral of Eq.~\ref{eq:conv-med} 
 over one oscillation cycle ($\theta=0\rightarrow \theta_{\text{A}}\rightarrow 0\rightarrow -\theta_{\text{A}}\rightarrow 0$) becomes zero, excluding directed motion.

\subsection{Sd-particles in a ring cluster} 

\begin{figure*}
\centerline{\includegraphics{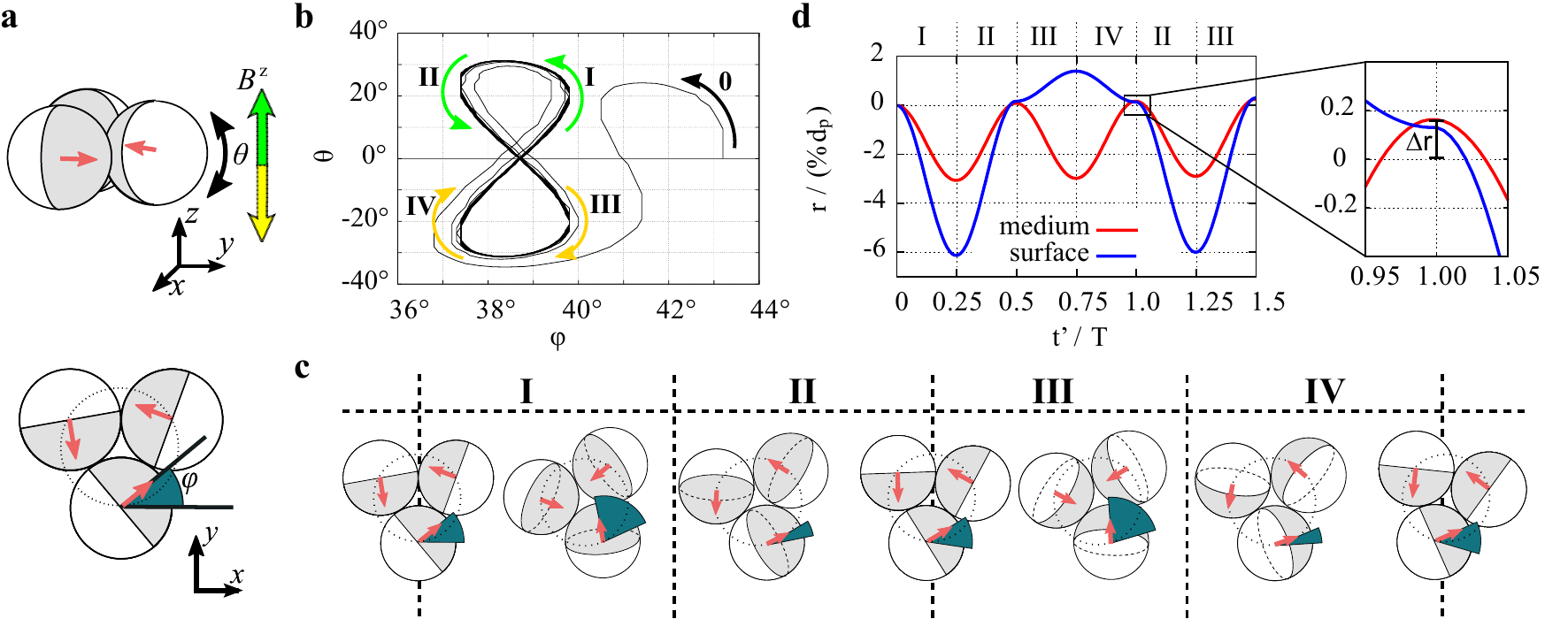}}
\caption{Rotation of a ring of sd-particles. (\textbf{a}) Triple ring (side and top view), and orientation angles $\varphi$ and $\theta$ of sd-particles. For better visualization, the hemisphere that contains the dipole (red arrow) is colored in grey.  The dotted ring indicates the orbit of motion spanned by the particle centers. (\textbf{b}) Numerically calculated trajectory in $\theta-\varphi$ coordinates of an sd-particle ($\xi=0.55$) in a triple ring under an oscillating field ($B_0^{\text{z}}=6\,\frac{\mu_0 m_{\text{p}}}{4\pi d_{\text{p}}^3}$). The arrows  indicate the time evolution. (\textbf{c}) Sketched top view of a triple ring during one movement cycle. For clarity,    the drawn variations in $\varphi$ (dark green segments) are much larger than the ones plotted in (b). (\textbf{d}) Orbital translation $r$ versus time $t'$ of an sd-particle in a triple ring that is located in an isotropic medium or on a surface ($c_1=2$ and $c_2=-0.4$). The time $t'$ is measured in period $T$. 
}
\label{graph:sk-rot}
\end{figure*}

Next, we transfer the considerations for a single sd-particle to sd-particles that form a two-dimensional cluster in the plane $z=0$. We assume that within the cluster the particles are free to rotate in response to magnetic stimuli but that the relative distances between all particles are kept constant during field exposure. The simplest case to study interacting sd-particles is given if they form a ring. 
Then, ideally all particles have identical boundary conditions and behave uniformly. As an example, a ring of three particles (triple ring, Fig. \ref{graph:sk-rot}\,a) will be studied. 
 
In analogy to a single sd-particle, rotation by $\mathrm{d} \theta$ displaces each sd-particle by $\mathrm{d} r$ and $\mathrm{d} z$. However, since the particles are constrained to a circle by the magnetic interaction (Fig.~\ref{graph:sk-rot}\,a) and move uniformly, the displacement $\mathrm{d} r$ is forced along an orbit spanned by the particle centers. As a consequence, the displacement of each sd-particle,  induced by the rotation $\mathrm{d} \theta$, is  hindered sterically by the presence of neighboring particles. The hindrance depends on the angle $\varphi$ that is enclosed by the tangent to that orbit and the in-plane component of  the dipole (Fig.~\ref{graph:sk-rot}\,a). $\mathrm{d}r$ becomes zero if $\varphi=90^{\circ}$, and it is completely unobstructed if  $\varphi=0^{\circ}$. Such an anisotropic hindrance is taken into account by introducing the reduction factor $c_{\text f}(\theta)=\cos \varphi(\theta)$. This hindrance indirectly affects also the displacement $\mathrm{d}z$.
          
The factor $c_{\text f}(\theta)$ suggests that the hindrance between interacting particles depends exclusively on $\varphi$. To determine $\varphi$ during the driven oscillations of sd-particles in a triple ring, we have studied the rotational (angular) motion of the particles separately from their translational motion.  
The rotational motion $\varphi(\theta(t))$ has been obtained by numerically solving the equation of rotation of the interacting sd-particles under an oscillating field $B^{z}$, which points  perpendicular to the plane of the ring cluster. 

The periodic variation of $B^z$ leads to an oscillatory behaviour of $\theta (t)$. Additionally, we have observed that the variation of $\theta$ involves a periodic change of $\varphi$  for interacting sd-particles in a ring. 
The trajectories  of the dipoles in $\theta$\,-\,$\varphi$  coordinates are double loops (Fig.~\ref{graph:sk-rot}\,b).  Starting from the field-free equilibrium state (0), the dipoles perform a short transient oscillation until they reach a steady-state oscillation along the path I\,$\rightarrow$\,IV.  
The loop implies that the dipoles obtain larger angles $\varphi$ while $|\theta|$ increases (I/III) than during the part of the trajectory in which $|\theta|$ decreases (II/IV) (Fig.~\ref{graph:sk-rot}\,c). Such a loop trajectory $\varphi(\theta)$ breaks time reversibility since the sense of direction of the loop is uniquely defined.  Inserting $\varphi(\theta)$ into the reduction term $c_{\text f}(\theta)=\cos\varphi(\theta)$ gives rise to a time-irreversible  translation of the sd-particles along the orbit. During increase of $|\theta|$, the displacement $\mathrm{d}r$ is hindered more strongly than during decrease of $|\theta|$. Based on that a non-reciprocal displacement $\mathrm{d}r$ of the sd-particles in the triple ring can be predicted, which leads to an effective rotation of the whole ring  (Fig.~\ref{graph:sk-rot}\,c).  



Inserting the term $c(\theta)=c_{\text f}(\theta)=\cos\varphi(\theta)$ 
into Eq.~\ref{eq:conv-med} gives the infinitesimal displacement of a particle  in a ring located in an isotropic environment. 
The time dependence of the displacement $\mathrm{d}r(t)$ can be obtained by inserting analytic expressions for $\varphi(\theta(t))$ and $\theta(t)$. Fitting the numerical data (Fig.~\ref{graph:sk-rot}\,b) to Lissajous curves (see SI Appendix) yields suitable approximate functional terms. The net translation is obtained by integration, $r(t')=\int_0^{t'} \mathrm{d}r(t)$ (Fig.~\ref{graph:sk-rot}\,d). 


For a ring in an isotropic medium, the sd-particles periodically translate back and forth twice per movement cycle (Fig.~\ref{graph:sk-rot}\,d) similar to a single particle (case ii, Fig.~\ref{pics:sk-trans-sd}). In addition, they gain an overall distance $\Delta r$ traveled after each movement cycle I\,$\rightarrow$\,IV. This is a very small effect since $r$ is in the range of  a few $\% \, d_{\text p}$, and $\Delta r$ is less than one tenth of that range.
 
 The translation $r(t)$ during one cycle can be detailed as follows.
During increase of $|\theta|$ (I/III), the particle translates  with its magnetic side ahead (Fig.~\ref{graph:sk-rot}\,c),  defined as back motion. Forth translation with the non-magnetic side ahead occurs during sections II and IV. 
As derived from the rotational trajectory (Fig.~\ref{graph:sk-rot}\,b), the factor $\cos \varphi(\theta)$ during I/III is smaller than during II/IV. Therefore,  $-\mathrm{d}r(\theta_{\rm I})=-\mathrm{d}r(\theta_{\rm III})<\mathrm{d}r(\theta_{\rm II})=\mathrm{d}r(\theta_{\rm IV})$.   The integration of $\mathrm{d}r$ over one field cycle is, thus, positive. The particles in a ring  perform a net translation along the orbit with the non-magnetic side facing forward. Finally, the uniform translation of all particles along the orbit gives an effective rotation of the whole ring in the plane of the particle centers. 
 
In addition to the translation $\mathrm{d}r$ in the plane of the triple ring, also the motion $\mathrm{d}z$ perpendicular to the ring has to be considered. Collectively, the particles move up and down, resulting in an oscillating up-and-down translation of the ring. Due to the broken symmetry caused by $\cos \varphi(\theta)$, the trajectory in $z$-direction is not reversible either. However, the vertical component $z$ that is gained during  $\mathrm{d}\theta>0$ is reversed during $\mathrm{d}\theta<0$, and the ring gains no overall distance along $z$.

For a ring that is located on a surface, the translation $r(t)$ has been obtained analogously (Fig.~\ref{graph:sk-rot}\,d). In this case, the factor  $c(\theta)=c_{\text f}(\theta)\cdot c_{r}(\theta)$ has been applied in Eq.~\ref{eq:conv-med}\,a. The translational behavior differs qualitatively from the one in an isotropic medium if a value $c_{r,2}<0$ is applied, i.e. if the surface drag dominates over the effect of the dipole shift on the translation. Then, negative values of $c_{r,2}$ reverse the sign of the displacements in sections III and IV. Thus, forth displacement takes place during II/III and back displacement occurs during I/IV (Fig.~\ref{graph:sk-rot}\,d). 
 During I/II, where $c_{r,1}>0$ is effective, the translation $r(t)$ looks similar to the one without surface. Further, since on the surface $|c_1|>|c_2|$, $\mathrm{d}r(\theta_{\rm I})+\mathrm{d}r(\theta_{\rm II})<\mathrm{d}r(\theta_{\rm III})+\mathrm{d}r(\theta_{\rm IV})$ holds.
 Also in this scenario the integrated forth displacement  $r$ (facing the non-magnetic side) is larger than the back displacement (magnetic side) during one field cycle. In the example presented in Fig.~\ref{graph:sk-rot}\,d, the ratio $c_2/c_1=-0.2$ has been applied. This ratio is also visible in the curve $r(t)$  by the ratio between the amplitude during I/II and the one during III/IV. 


The presented  actuation concept is, of course, not limited to the triple cluster, which is, however, the smallest entity showing this effect. The concept can be applied to other clusters, too, as will be shown next by experimental studies.

 \subsection{Actuated clusters of capped particles}
 \label{ssec:rot-caps}
 
The proposed actuation of interacting sd-particles under oscillating fields can indeed be found in an experimental system. We use particles with a diameter of $d_{\text{p}}=4.54\,\mu$m that are equipped with  a hemispherical magnetic coating, so-called capped or Janus particles.  Due to the hemispherical coating, the magnetic center of mass is shifted away from the particle center.  A ferromagnetic coating with perpendicular magnetic anisotropy provides a stray field with dipolar characteristic \cite{Alb05}, and the net dipole points along the Janus director.  For these particles, an effective dipole shift of about $\xi=0.6$ has been estimated from comparison with  simulations  \cite{Kan11, Ste16b}. 

In solution, the particles sediment on a substrate and assemble in a great variety of open and compact two-dimensional structures with diverse magnetic order \cite{Ste16a,Ste16b,Bar08a}. The particle orientation can be visualized  by transmission light microscopy via the optical contrast between the transparent, uncoated and the intransparent, coated hemisphere. In contrast to the sd-particles with a single point dipole, the capped particles provide an extended magnetization distribution. This, however, leads only to a slight modification of the dipole interaction potential  \cite{Ste16b}. Quantitative consequences on the motion will be pointed out accordingly.  
 
\begin{figure}
\begin{center}
\includegraphics{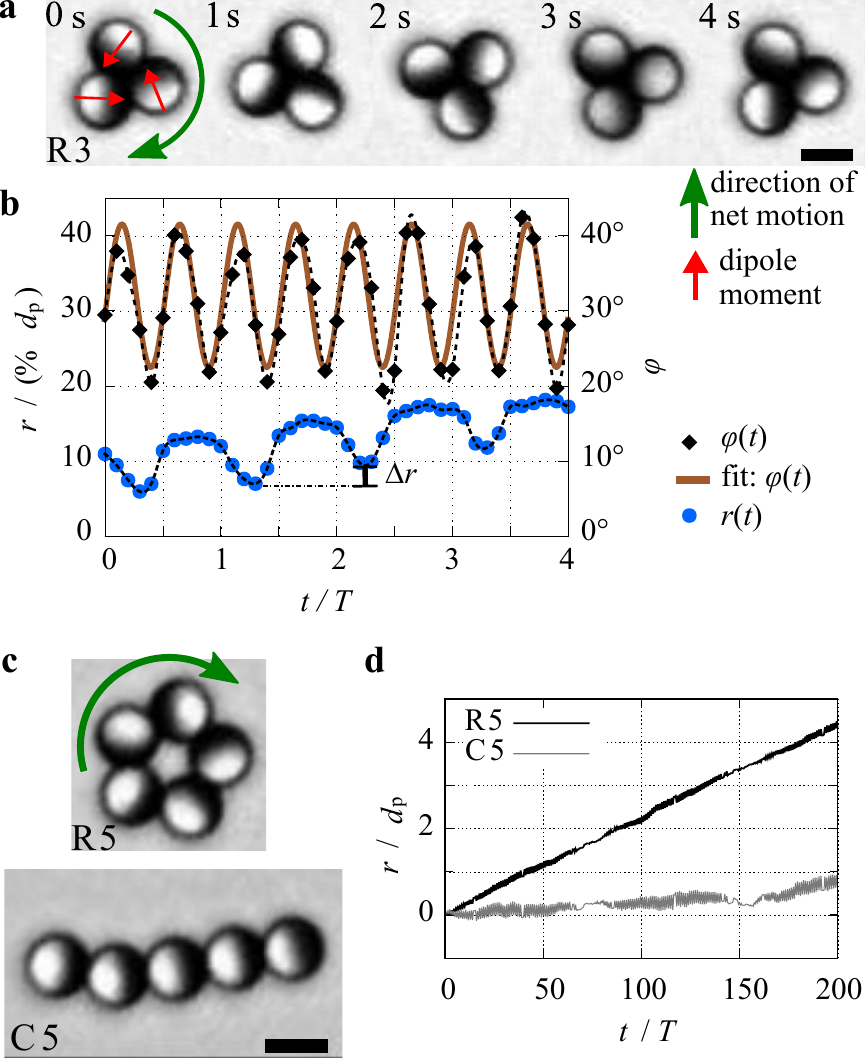}
\caption{Actuation of colloidal clusters. (\textbf{a}) Image sequence of the rotation of a self-assembled triple ring (R3) under an oscillating field. 
 (\textbf{b}) Orientation $\varphi(t)$, the sine fit curve (solid), and the orbital translation $r(t)$  of a capped particle in  R3 during four periods $T$ ($B_0^{\text{z}}=0.34$\,mT, $\omega_B/2\pi= 2$\,Hz at 20\,fps). Dotted lines are spline curves through the data points.  (\textbf{c}) Microscopy images of a five-particle ring (R5) and a linear chain (C5) that form   under weak oscillating fields. (\textbf{d}) $r(t)$ of a particle in R5 and C5 ($B_0^{\text{z}} = 1.22$\,mT, $\omega_B/2\pi= 10$\,Hz).  The superimposed beat results from the stroboscopic recording with a digital camera (20\,fps).  (Scale bars: 5\,$\mu$m)}
\label{graph:3-4ring}
\end{center}
\end{figure}

 Here, the actuation of selected structures  under oscillating fields will be presented. Spontaneously, the capped particles self-assemble into triple rings (R3). Under oscillating fields $B^{\text{z}}$, one can observe that the ring rotates (Fig.~\ref{graph:3-4ring}\,a). In agreement with the theoretical predictions, the uncapped, non-magnetic hemisphere always faces towards the rotation direction. The translations $r(t)$ and the orientations $\varphi(t)$ of the capped particles in R3 have been 
measured with image analysis (Fig.~\ref{graph:3-4ring}\,b). Their trends coincide qualitatively with  numerical findings for sd-particles in a ring on the substrate (Fig.~\ref{graph:sk-rot}\,d). During each cycle, $r(t)$ exhibits a motion sequence of  fast forth, slow forth, slow back, and fast back translation. In analogy to the simulation, the ratio between the amplitudes during the section of slow and fast translation in the experimental curve (Fig.~\ref{graph:3-4ring}\,b) gives a measure for the ratio between the conversion factors, suggesting a value of  $c_2/c_1\approx -0.2$. The mean relative increase $\Delta r/d_{\rm p}$ per cycle is $1.8\,\%$. 
This is one order of magnitude larger than in the simulation (Fig.~\ref{graph:sk-rot}\,b). The difference results from the modified dipole interaction between the capped particles, which will be confirmed next.  The capped particles perform radial oscillations  $\varphi(t)$ twice per field cycle T (Fig.~\ref{graph:3-4ring}\,b), which also coincides with the sd-particles. Fitting $\varphi(t)$ to a sine function gives  a mean value of $\varphi=32^{\circ}$ and an oscillation amplitude  of $10^{\circ}$. This radial amplitude of capped particles is about 8 times as large as for sd-particles, assuming same polar amplitudes in $\theta$. Therefore, also the double loop trajectory of capped particles is broadened by this factor. Since the area enclosed by the double loop is proportional to the propulsion efficiency $\Delta r/d_{\rm p}$ , the experimental value of $\Delta r/d_{\rm p}$ must be about one order of magnitude larger than in the simulation.

The theoretically proposed vertical displacement of the particle centers are not visible in the  vertical projections of the clusters. However, indirect evidence is given by the experimentally obtained value of $|c_2/c_1|\neq 1$, which implies that the translational friction differs between $\theta<0$ and $\theta>0$. In order to exclude an influence based on two chemically different surfaces of the Janus particles, we have examined two sets of particles, one with a Pd layer and one with a silicon oxide layer on top of the magnetic film. Under same experimental conditions, triple rings of both types of particles exhibit the same angular speed, thus, surface effects can be excluded. 
 

Next, we have tested whether the actuation concept also holds for other clusters.
Due to the magnetic shift, capped particles (and also sd-particles) do not form larger rings  spontaneously. 
However, we have shown earlier that closed rings and linear chains (R5 and C5 in Fig.~\ref{graph:3-4ring}\,c) can form if weak, low-frequency oscillating fields are applied \cite{Ste16a}. The presence of these clusters  is, thus, linked directly to any motion of these clusters under oscillating fields. We have observed that also R5 rotates under oscillating fields, leading to a significant slope in $r(t)$ (Fig.~\ref{graph:3-4ring}\,d). 
 In contrast, under the same environmental conditions, chains (C5)  only perform periodic back and forth displacement along the chain direction with almost no net translation.  
As  reported previously, the loop trajectory (\ref{graph:sk-rot}\,b), which  breaks time-reversal symmetry, vanishes for a collinear orientation of the dipolar particles \cite{Ste16a}. The particles  perform reciprocal oscillations $\theta(t)$ without variation in $\varphi$. This implies that clusters of collinear sd-particles cannot move under oscillating fields.
  The small, but yet visible, slope in the trajectory of C5  (Fig.~\ref{graph:3-4ring}\,d) suggests that  the particles in a chain are not perfectly collinear at all times. This might be caused by thermally induced bending (deformation) of the chain. 

\begin{figure}[t]
\includegraphics{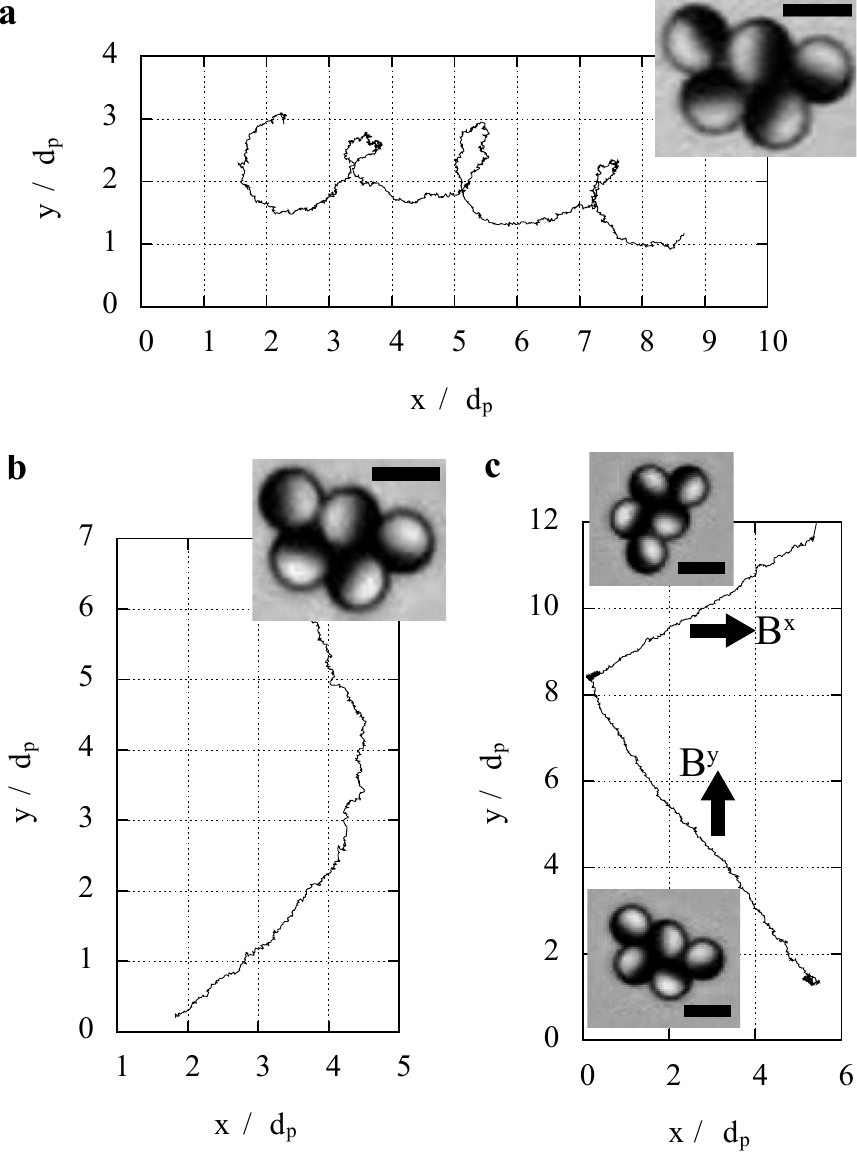}
\caption{Trajectory of the center of mass of two compact clusters  (insets) (\textbf{a,b})  under  an oscillating field with $B^{\text{z}} = 0.53$\,mT, $\omega_B/2\pi = 5$\,Hz  (a: 88\,s; b: 251\,s)  and (\textbf{c}) under an oscillating field ($B^{\text{z}} = 1.13$\,mT, $\omega_B/2\pi = 20$\,Hz) and a constant in-plane field ($B^{\text{x/y}} = 0.03$\,mT). The orientation of $B^{\text{x/y}}$ is switched from the $y$-axis to the $x$-axis during the recording, leading to a change of the translation direction. (Scale bars: 5\,$\mu$m)}
\label{graph:rot-trans}
\end{figure}
So far, the actuation of interacting sd-particles  has been demonstrated for ring clusters to take advantage of symmetric boundary conditions. Of course, the presented concept also holds if the assembled cluster is not rotationally symmetric.  
Since the motion of a cluster arises from the sum of the displacements $\mathrm{d}r$ of the constituting particles, the lack of rotational symmetry can result in more diverse motion. 
 As an example, two compact clusters consisting of five particles (Fig.~\ref{graph:rot-trans}\,a, b) have been examined under oscillating fields. The capped particles can self-assemble into compact clusters with various magnetic configurations as a result of the magnetic shift (SI) \cite{Kan11,Kli13}. Here, we present two stable clusters (Fig.~\ref{graph:rot-trans}\,a, b) that exhibit the same spatial shape but slightly different magnetic configurations. Specifically, they differ in the azimuthal orientation of only one particle (most left one in the insets in Fig.~\ref{graph:rot-trans}) while all other four particles are similar. 
 Interestingly, this small difference in the magnetic configuration leads to a drastic difference in the trajectory of the clusters when actuated by oscillating fields (Fig.~\ref{graph:rot-trans}\,a, b). One of the clusters (a) performs a screw-like motion while the other (b) exhibits an almost linear path. Considering that each particle in a cluster contributes with  $\mathrm{d}r$ along the in-plane orientation of its cap, the net translation arises approximately from the sum of the dipole vectors. In cluster (a), the dipole vectors of the particles form a bent ring, which leads to the observed helical motion. In cluster (b), the vectorial sum of the cap orientations gives a significant cluster moment, resulting in  directed translation of the cluster.  

The net cluster moment additionally can be used to orient the cluster in fields parallel to the cluster plane, provided that the field intensity is too low to alter the cluster structure. This enables a control over the translation direction (Fig.~\ref{graph:rot-trans}\,c). After orienting the cluster along a weak parallel field, it performs directed translation under an additional oscillating field. Switching the in-plane field orientation by $90^{\circ}$ results in a change of the translation direction. A noteworthy feature in the example presented in Fig.~\ref{graph:rot-trans}\,c is that the translation path is not parallel to the direction of the applied field, because the net magnetic orientation of this cluster does not coincide with its effective translation direction.

\section{Discussion}

At low Reynolds numbers, locomotion under an oscillating drive is possible only if the system provides flexibility. 
Here, we have shown that this can be realized even in the absence of structural body deformations, a scenario that is impossible within the usually considered actuation concept of hydrodynamic drag. The newly introduced propulsion strategy  requires two ingredients. First, an object must be built from multiple spheres that can rotate internally in response to a field. Second, the spheres must exhibit an anisotropic magnetization distribution. Under oscillating fields, the interaction between these spheres activates torsional oscillation perpendicular to the field direction, which breaks time reversal symmetry. 
While this strategy does not rely on hydrodynamic drag, the presence of drag close to a surface  modifies the motion. We propose that  this actuation concept can be applied also to other magnetically anisotropic particles, which are currently widely discussed for complex self-assembly \cite{Tie14,Che09,Yen16,Sac12,Abr13,Yan12}.

For practical applications, the presented actuation concepts combines three essential benefits. First and most striking, the overall motion can take different paths depending on the  magnetic configuration of a cluster. Second, the translation direction can be controlled. Third, from the equation of motion  (Eq.~\ref{eq:erot-coup}) it can be deduced that the speed of motion is adjustable by varying the amplitude of the oscillating field. 

 Ring clusters perform rotation under oscillating fields due to their rotational symmetry. Other, asymmetric clusters perform directed translation as demonstrated with magnetically capped particles. The motion crucially depends on the position and orientation of each constituting particle in the cluster.  This suggests that, beyond the presented examples, by proper choice of asymmetric particles and assembly of a cluster  any combination of rotation and translation can be realized. For a rigorous theoretical description of a translating cluster with asymmetric shape, additionally, the anisotropic drag must be taken into consideration, which provides a complex problem on its own \cite{Kuem13,Wen14}.
Further, a translating cluster was shown to exhibit a residual magnetic moment, which can be used to steer the cluster orientation with a weak parallel field. A superimposed oscillating field leads to orientation-controlled directed translation. Thus, the actuation with a single field component advantageously gives two remaining field components free for controlling the translation direction. Finally, the field amplitude directly controls the actuation speed. With increasing fields,  the oscillation amplitude $\theta_A$ increases. Since $\theta_A$ sets the integration bounds for the displacement $\Delta r$ per field cycle (Eq.~\ref{eq:conv-med}), the field amplitude determines $\Delta r$. The presented clusters move with 1-5\,$\mu$m/s, and, thus, the actuation efficiency is comparable to those of swimmers under anisotropic drag \cite{Bec16}.

In conclusion, the presented results show that internal rotations in multi-component objects can play a crucial role for directed locomotion in overdamped systems. This provides new actuation strategies that complement established concepts based on anisotropic drag. Considering the simplicity of the oscillating field pattern that drives the motion, the question remains whether similar actuation can be found elsewhere, e.g., in animate systems.
In addition to anisotropic potentials, also  entropic interactions between components with anisotropic shape \cite{Glo07} might enable similar propulsion mechanisms based on the presented strategy of internal rotations.


%
%

\section{Methods}
\subsection{Numerical Methods} 
The rotational motion of interacting sd-particles in a ring cluster exposed to an oscillating field $\mathbf B^{\rm{z}}=\mathbf B^{\rm{z}}_0\sin (\omega t)$ has been calculated by numerically solving the equation of rotation. In the simulation, the particles are spatially fixed at their center. This assumption is reasonable when studying the rotation of the particles in the reference system of a cluster that does not change its spatial shape. Assuming sd-particles  with dipole moment $\textbf{m}$ (unit vector $\hat{\textbf{m}}$) and stray field  $\textbf{B}^{\rm{p}}$, the equation of rotation is obtained by balancing the rotational drag, ${\tau}^{\rm d}$, with the magnetic torques between the sd-particles and between sd-particles and the field, ${\tau}^{\rm B}$. Inertial forces have been neglected on account of the dominating drag forces.  Due to the dipole shift, besides the aligning torque ${\tau}^{\rm p}=\textbf{m} \times \textbf{B}^{\rm{p}}$, also the gradient force $(\textbf{m} \cdot \nabla) \textbf{B}^{\rm{p}}$ between the sd-particles is relevant. 
The latter converts into an effective torque, ${\tau}^{ F}$, such that the equation of rotation is given by
\begin{eqnarray}
{\tau}^{\rm d}&=& {\tau}^{\rm p}+ {\tau}^{F} + {\tau}^{\rm B}\\
f_{\rm{r}} \dot{\Theta} (t)&=& \textbf{m} \times \textbf{B}^{\rm{p}} + \xi \hat{\textbf{m}} \times  ((\textbf{m} \cdot \nabla)\textbf{B}^{\rm{p}}) + \textbf{m} \times \textbf{B}^{\rm{z}}
\label{eq:erot-coup}
\end{eqnarray}
$\dot{\Theta}$ is the angular velocity of an object with rotational friction coefficient $f_{\rm{r}}$ and orientation  $\Theta=(\theta,\,\varphi)$ (Fig.~\ref{graph:sk-rot}\,a). An ensemble of $n$ particles gives a system of $n$ coupled  equations of rotation (Eq.~\ref{eq:erot-coup}), which are solved iteratively in a self-consistent cycle. From the state at time $t$, the magnetic moment $\mathbf{m}_i$ of particle $i$ at time step $t+1$ is obtained by
\begin{equation}
\hat{\textbf{m}}_i^{(t+1)}={\rm{norm}}\left[\hat{\textbf{m}}_i^{(t)} + \frac{1}{f_{\rm{r}}}(\tau^p + {\tau}^{F} + {m}_i\times ({B}^{z}_0 \, \cos(\omega t)) )\right]
\label{eq:rot-mot}
\end{equation}
The function 'norm$[\cdot]$' scales the updated magnetic moments $\mathbf{m}^{(t+1)}$ to retain the initial magnitude \textit{m} of the dipoles. In the calculation, a rotational friction coefficient of $f_{\rm r}=100\,\frac{\mu_0 m^2}{32\pi r_{\rm p}^3}$ has been applied. It has the dimension of energy since the numerical time steps $\Delta t=(t+1)-t$ are dimensionless and  set to 1. The angular frequency $\omega$ is also a dimensionless quantity and is set to 0.05.
\subsection{Experimental Details} 
The particle preparation and experimental setup has been described in detail in reference \cite{Ste16b}. In short, silica spheres with a radius of $r_{\rm p} = (2.27 \pm 0.23)\,\mu$m have been coated on one hemisphere with a magnetic thin film of $[{\rm Co}(0.28\,{\rm nm})/{\rm Pd}(0.9\,{\rm nm})]_8$ \cite{Alb05}. The magnetic thin film causes mass imbalance in the subatomic range (SI) and, thus, can be neglected here.  This film is known to exhibit  strong perpendicular magnetic anisotropy \cite{Car85}; the magnetic orientation points perpendicular to the film plane. When depositing the film on a sphere, the magnetic anisotropy follows the curvature of the particle surface. After magnetic saturation, the particle becomes magnetically single domain and obtains a radially symmetric anisotropy distribution \cite{Alb05}. This leads to a magnetic stray field with dipolar characteristic, and the net dipole moment points along the Janus director. To test the influence of the chemically distinct surfaces, one set of particles has additionally been coated with $\text{Si}_3\text{N}_4$ on top of the metal film. This top layer oxidizes and becomes amorphous silicon oxide, which is chemically similar to the uncoated silica hemisphere.  

 A suspension of the coated particles in distilled water is studied via transmission  light microscopy. Due to the density mismatch the particles sediment on the ground of the sample cell, providing a two-dimensional particle system. An electromagnetic coil is mounted above  the sample cell, providing perpendicular, low-frequency fields. An additional set of two pairs of coils attached beneath the sample provide constant fields to orient translating clusters. The microscopy recordings of the moving clusters have been analyzed by optical image analysis. Using the open-source software ImageJ, a customized detection algorithm based on Hough transformation was built and employed.  

\subsection{Acknowledgments}

Financial support by the German Research Foundation (DFG, Grant Nos. ER 341/9-1, AL618/11-1, and FOR 1713 GE 1202/9-1) is gratefully acknowledged. We are thankful to Bjoergvin Hj\"orvarsson for pointing to the rotating clusters in the microscopy recordings. We thank Henrique Moyses and  Jeremy Palacci for fruitful discussions.
\subsection{Author contributions}
G.S. and A.E. conceived the experiments, G.S. and S.G. conceived the simulations. M.A. devised the routine for particle coating and D.N. prepared the particles.  G.S. conducted the experiments and the simulations and interpreted the results. M.S. gave guidance and advice. G.S., A.E. and S.G. wrote the paper. All authors reviewed the manuscript.
\subsection{Correspondence}
Correspondence and requests for materials should be addressed to G.S. (gabi.steinbach.de@gmail.com) and to A.E. (a.erbe@hzdr.de).
%

\end{document}